\documentclass[amssymb,amsmath,prl,floatfix,onecolumn,nofootinbib]{revtex4}
\usepackage{graphicx}
\usepackage[section] {placeins}
\usepackage{color}
\usepackage{graphicx}
\usepackage{lastpage}
\usepackage{amssymb}
\usepackage{amsmath}
\newcommand{\ket}[1]{\left|#1\right\rangle}

\newcounter{para}

\begin{document}
\title{High-speed quantum networking by ship}

\author{Simon J. Devitt$^{1}$, Andrew D. Greentree$^2$, Ashley M. Stephens$^3$ and 
Rodney Van Meter$^4$}\affiliation{$^1$Center for Emergent Matter Science, RIKEN, Wakoshi, Saitama 315-0198, Japan.} \email{simon.devitt@riken.jp}
\affiliation{$^2$Australian Research Council Centre of Excellence for Nanoscale BioPhotonics, and Chemical and Quantum Physics, School of Science, RMIT University, Melbourne 3001, Australia}
\affiliation{$^3$National Institute of Informatics, 2-1-2 Hitotsubashi, Chiyoda-ku, Tokyo 101-8430, Japan.}
\affiliation{$^4$Faculty of Environment and Information Studies, Keio University, Fujisawa, Kanagawa 252-0882, Japan.}

\date{\today}
\begin{abstract}
Networked entanglement is an essential component for a plethora of quantum computation 
and communication protocols.
  Direct transmission of quantum signals over long distances is prevented by fibre attenuation and the
  no-cloning theorem, motivating the development of quantum repeaters, 
  designed to purify entanglement, extending its range. Quantum repeaters
  have been demonstrated over short distances, but error-corrected, global repeater networks with
  high bandwidth require new technology.  Here we show that error 
  corrected quantum memories installed in cargo containers and carried by
  ship can provide a flexible connection between local networks, enabling low-latency,
  high-fidelity quantum communication across global distances at higher bandwidths than 
  previously proposed.  With demonstrations of 
  technology with sufficient fidelity to enable topological error-correction, implementation of the
  quantum memories is within reach, and bandwidth increases with improvements in
  fabrication. Our approach to quantum networking avoids technological restrictions of
  repeater deployment, providing an alternate path to a worldwide Quantum Internet.
  \end{abstract}
\maketitle

%{\color{blue}209}

%\section{Significance Statement}
% ?Never underestimate the bandwidth of a station wagon full of tapes
%hurtling down the highway?. This famous quote encapsulated the power of a lowtech
%approach to information transfer: sneakernet. Even today, sneakernet has higher
%bandwidth, lower cost and greater security than optical fibre networks. But optical fibre has
%the advantage of low latency: we get information when we want it, limited by the speed of
%light and routers. We introduce the quantum analogue of sneakernet: quantum sneakernet.
%This approach can be used to construct a worldwide Quantum Internet with performance
%vastly superior to traditional approaches. Unlike classical sneakernets our approach
%distributes entanglement rather than information, allowing quantum information transfer at
%the speed of sending a classical signal. With quantum sneakernet as the ?long-haul?
%distribution channel for entanglement, we can ask serious questions about how to build a
%global quantum internet. 

%\section{Introduction}

Quantum communication will strengthen the security of cryptographic systems and decision-making
  algorithms\cite{BB84,E91,OH05,BR03}, support secure client-server quantum computation\cite{BFK09}, and
  improve the sensitivity of scientific instruments\cite{GJC12,BRS07,JADW00}.  An effective global 
  quantum network is ultimately required to support such applications\cite{K08+,LSWKSY04,V14} 
  and significant research has been conducted to realise such networks\cite{BDCZ98,F10,JTNMML09,LBSB13}.  The most common method is the transmission of encoded 
  optical signals along traditional fibre connections between more localised networks.

Photons are traditionally proposed for the establishment of quantum
entanglement between stationary quantum systems over moderate
distances. Over longer distances, entanglement purification and
entanglement swapping connecting a path comprised of shorter links
will mitigate the exponential attenuation loss, the no-cloning theorem\cite{WZ82} and effects of
imperfect devices \cite{BDCZ98}. The repeat-until-success nature of
these techniques allows the rate of entanglement generation to
decrease polynomially with increasing distance, with the fidelity of
entanglement limited by the accuracy of the quantum gates operating in
the repeaters.  While experimental demonstration of short-range quantum communications has 
been effective\cite{R12,HI14}, long range repeater networks require the
incorporation of fault-tolerant error correction methods, and numerous
designs have been proposed
\cite{F10,JTNMML09,LBSB13,MK14,ATL13,DMN08,MHSDN10}.  These designs do not
offer bandwidth higher than about a MHz and require dense repeater
arrays.  A global network constructed in this way would require
high-power, low-temperature quantum devices with active control
deployed in very hostile environments.  While the deployment of satellite technology may 
mitigate this problem \cite{BBMHJS14}, a fully error corrected, global system 
has still not been developed.   Consequently, 
no known technology meets the stringent requirements for global deployment. In this work we present an alternative method 
that could be used to augment networks based on traditional repeaters and satellite technology.  This approach can possibly solve 
crucial issues related to a global network in a way that remains compatible with other well known networking schemes.

%More sophisticated repeaters incorporating quantum error correction may achieve better
%scaling with distance and retain higher fidelity \cite{F10,JTNMML09,LBSB13}, but a global network of error-corrected
%repeaters will require high-power, low-temperature quantum devices with active control in dense deployment along
%undersea cables. No known technology meets this requirement.

Despite the capabilities of modern classical networks, it is still routine to transfer classical information
stored in removable media, an approach known as \emph{sneakernet}. Here we adapt this approach to quantum
information, introducing a new network mechanism for the establishment of quantum entanglement over long
distances based on the transport of error-corrected quantum memories \cite{DKLP02}. Long-distance information transport
involves significant latency, but establishing entanglement involves no exchange of users' data, meaning that
this latency is irrelevant to network users. Indeed the fact that entanglement distribution can occur without the exchange of classical information leads immediately to Peres' well-known `paradox' of delayed-choice entanglement swapping \cite{Per00,MZK+12}.  The fact that classical information need not be exchanged to distribute entanglement means that quantum networks can be effectively zero latency, although of course any protocol that uses the distributed entanglement,  for example, for quantum teleportation \cite{BBCJPW93}, will be limited by the classical one-way information transfer speed. Quantum memories may be
transported to locations where entanglement is required or to intermediate locations to facilitate
entanglement swapping between traditional quantum repeater networks, enabling a complete network structure
without the full deployment of physical links. Because transoceanic communication presents a particular
challenge for quantum networking, we focus on the establishment of entanglement by ship.

This approach of creating a quantum sneakernet through physical transport of actively error corrected quantum memories will require the fabrication of more physical qubits in the overall network, but has the possibility of significantly mitigating issues related to deployment of a quantum network.  The most effective quantum repeater designs for entanglement distribution (so called 3rd generation repeaters) employ full 
error correction protocols and consequently require close separation (within of order 10km) 
due to the need for low loss connections \cite{F10,LBSB13}.  Each repeater unit is effectively a mini-quantum computer, requiring infrastructure technologies such as cryogenic cooling and vacuum systems, extensive classical control (for both qubit manipulations and error-correction/network operations) and non-trivial power requirements.  The deployment of such devices, every 10km, across an ocean constitutes an enormous engineering challenge.  Conversely, a sneakernet approach allows us to convert this significant hurdle in deployment to the challenge of making mini-quantum computers amenable to physical transport.  This allows us to design memory units that can be serviced (or replaced) at regular intervals with minimal impact on the overall network or incremental upgrades to the entire system as technology becomes denser and/or cheaper.  This sneakernet method of quantum networking provides an alternate pathway that could augment traditional repeater and satellite based systems where the issue of network deployment and servicing in hostile environments becomes extremely hard or where entanglement nodes are required on mobile platforms not accessible by a traditional repeater node.

%\section{Results}

Our approach requires quantum memories with an effective coherence time of months, sufficient for transport
along any traditional shipping channel. Because an error-corrected quantum memory is based on the same system
architecture as a large-scale quantum computer, technology currently in development will satisfy our needs
\cite{JMFMKLY10,YJG10,N14}. In particular, implementations of qubits based on several physical
systems---including superconducting circuits and trapped ions---are nearing the accuracy threshold required
for topological error correction to become effective \cite{B14}. Once this threshold is exceeded, it will be
possible to arbitrarily lengthen the effective coherence time of an error-corrected quantum memory with a
poly-logarithmic qubit overhead. However, this new mechanism also requires quantum memories that are
compatible with storage and transport in a shipping container, including a stable power source, ultra-high
vacuum or refrigeration systems to maintain appropriate operating conditions, and classical-control
infrastructure to perform error correction and decoding \cite{D14}. As yet, there are no implementations
designed with this degree of portability in mind.  However, keeping in mind these 
engineering constraints, we will, for concreteness, consider a potential candidate system: an 
implementation based on negatively charged nitrogen vacancy (NV$^-$) centres in diamond, which may be
integrated in dense arrays and are optically accessible at a temperature of 4 K\cite{N14,DSSB13}.

The active nature of the quantum memory is designed to correct traditional errors arising from environmental decoherence and imperfect control.  However, mobile quantum memories may be subject to additional errors arising from physical movement.  These additional errors can arise from changes in, for example, the local magnetic field of the earth or from motional instability of the actual qubits.  Most physical systems currently considered for large-scale quantum memories are immune from motional instability ether because they are etched circuits (superconductors, linear optics) or because they consist of atoms locked into a physical lattice (NV-diamond, silicon).  Changes in global fields are not expected to cause significant problems for physical transport of these devices because global fields will vary on time scales that are far slower than the internal 
error correction utilised by each unit.  Physical transport does introduce "catastrophic" failure channels (for example, physically losing the memory unit), but these failure channels do not impact the analysis presented in this work.

Each quantum memory consists of a two-dimensional array of optical cavities with a mean linear spacing of 0.66 mm, each
containing a single NV$^-$ centre comprising a spin-half N$^{15}$ nucleus and a spin-one electron\cite{N14,DSSB13}. 
Each nuclear spin
represents a single qubit, and all operations (initialisation, readout, and interactions between neighbouring qubits)
are achieved by hyperfine coupling to the electron spins and dipole-induced transparency in an external optical
field. We assume that these operations are fixed to a 3.5 $\mu$s clock time and occur with an independent depolarising
error rate of 0.1\%, per gate \cite{N14}. Coherent control of spins in diamond has already been demonstrated with an error rate
below 1\%, indicating that this target may be achievable in the near future \cite{DB14}. Each quantum memory stores a
single logical qubit encoded in the surface code, although other codes may also be suitable. The error-correction
protocol involves the continuous execution of physical quantum circuits to determine the error syndrome. This
information undergoes local classical processing to detect and correct errors introduced by decoherence, coupling
inefficiency, and other sources, thereby preserving the state of the logical qubit. To determine the effective coherence
time of the quantum memories, we undertook numerical simulations of the error-correction protocol for small arrays. 
(see
the supplementary material). Extrapolating to large arrays, we find that in this system, a quantum memory of approximately 4200 qubits enables storage
of a logical qubit for approximately 40 days with a logical error rate of $10^{-10}$ [See the methods section].

\begin{figure*}
\begin{center}
\resizebox{160mm}{!}{\includegraphics{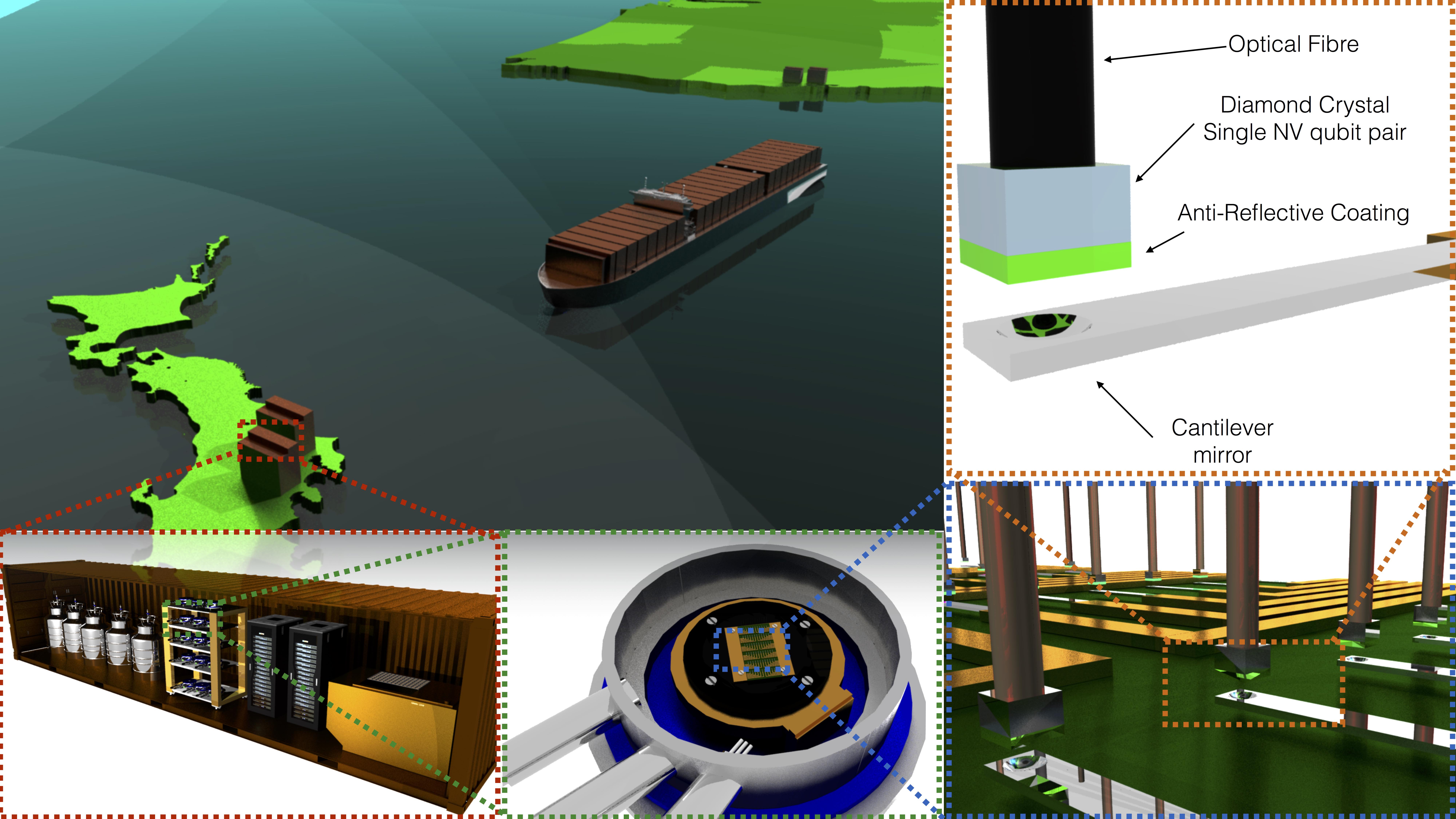}}
\end{center}
\vspace*{-15pt}
\caption{{\bf Physical transport protocol for a transpacific sneakernet using a single ship.}  The
  transpacific connection shows the location of memory stick units both on shore in the U.S. and Japan and in
  transit aboard a VLCS-class ship.  Each cargo container (memorystick) contains the actual memory units as
  well as any required control, cooling and power infrastructure.  Each memory unit (for the specific hardware
  model of optically connected NV$^-$ qubits \cite{N14}) consists of an array of diamond crystals within
  adjustable single sided cavities which are optically connected to perform individual qubit/qubit
  interactions. \cite{DSSB13}.}
\label{fig:transport}
\end{figure*}

Quantum memories are installed in Twenty-foot Equivalent Unit (TEU) containers, the standard shipping unit
with an internal volume of 40 m$^3$. We assume that 1 m$^3$ is occupied by quantum memories and the remaining
39 m$^3$ is reserved for power, refrigeration, and control infrastructure.  Each of these units is the quantum
equivalent of a memorystick.  More efficient protocols may be achievable under different assumptions, but for
simplicity we assume a single, dedicated Very Large Container Ship (VLCS)-class container ship with a capacity
of $10^4$ TEU. %, giving a total capacity of $10^7$ logical qubits per ship.
We consider a shipping channel between Japan and the United States, with freight terminals acting as primary
network nodes for traditional repeater networks deployed in each country, as shown in Figure
\ref{fig:transport}. Allowing for local transport and maintenance, this channel requires a one-way transport
time of 20 days [http://www.joc.com/sailings, "Journal of Commerce, sailing schedules", (2016) (Date of access:01/07/2014)].

Table I reports the effective transpacific bandwidth of the network mechanism. Our results compare quantum
memories based on NV$^-$ centres in diamond to quantum memories based on a range of other qubit
implementations \cite{YJG10,JMFMKLY10,N14,BKRB08,KHL14,HGFW06,M09,D09} illustrating the dependence of the
bandwidth on the underlying physical parameters. Each implementation involves a unique set of technological
challenges and our results are predicated on the development of portable systems incorporating
high-speed external interfaces to facilitate lattice-surgery operations between logical qubits. Nevertheless,
bandwidth in excess of 1 THz is feasible under realistic physical assumptions, exceeding even the fastest
proposals for traditional repeater networks. Our results assume a single container ship, but the total
bandwidth scales linearly with the total freight capacity, allowing for incremental investment in
infrastructure rather than the overhaul of thousands of kilometres of undersea cables. Furthermore, adding
network nodes involves transport to additional locations rather than investment in wider area infrastructure,
and can be done with minimal planning and only a few weeks lead time.

\begin{table*}
\begin{tabular}{ccccccc}
\hline
implementation 	& qubit pitch (m)		& gate time (s) 			& physical error rate 	& ($d$, $N$)& Memorystick capacity	& bandwidth (Hz)		\\
\hline
NV$^-$ (optical) 	& 6.6$\times$10$^{-4}$\cite{DSSB13}	&
3.5$\times$10$^{-6}$\cite{N14} 	& 1$\times$10$^{-3}$ & (33, 4225) & 12.7KEb 	& 7.3$\times$10$^{1}$ 	\\
trapped ions		& 1.5$\times$10$^{-3}$\cite{KHL14}	&
1.0$\times$10$^{-4}$\cite{BKRB08}	& 1$\times$10$^{-5}$ &(11, 441) & 32KEb	& 1.9$\times$10$^{2}$      \\
transmons			& 3.0$\times$10$^{-4}$\cite{D09}
& 4.0$\times$10$^{-8}$\cite{B14}	& 1$\times$10$^{-5}$&(13, 625) & 2.4MEb	& 1.4$\times$10$^{4}$   	\\
quantum dots		& 1.0$\times$10$^{-6}$\cite{JMFMKLY10} 	&
3.2$\times$10$^{-8}$\cite{JMFMKLY10}	& 1$\times$10$^{-3}$& (36, 5041) &2.8TEb	& 1.6$\times$10$^{10}$	\\
NV$^-$			& 3.0$\times$10$^{-7}$\cite{YJG10} 	&
1.0$\times$10$^{-3}$\cite{YJG10} 	& 1$\times$10$^{-3}$&(29, 3249) &200TEb	& 1.6$\times$10$^{12}$	\\
silicon			& 2.0$\times$10$^{-7}$\cite{M09}  	&
5.0$\times$10$^{-8}$\cite{HGFW06}	& 1$\times$10$^{-3}$ &(36, 5041) &350TEb	& 2.0$\times$10$^{12}$	\\
\hline
\end{tabular}
\label{tab:results}
\caption{{\bf Effective bandwidth of a transoceanic link.} Effective
  bandwidth achieved using a single VLCS-class container ship
  transporting error-corrected quantum memories between Japan and the
  United States, estimated for a range of qubit implementations for a
  fixed infidelity of $1-F = 10^{-10}$. For several implementations,
  bandwidth exceeds the fastest proposals for traditional repeater
  networks at far lower infidelities.  Memorystick capacities (in
  Entangled bits, or Ebits, Eb) are
  estimated as $1/(\sqrt{N}\times \text{pitch})^3$ for a memory unit
  of $N$ qubits given the operational gate time and error rate for a
  40 day storage time when utilising 1m$^3$ of space within each
  container.   Figures for 
  physical error rates are development targets for production-use hardware.  They are chosen assuming more experimentally
  mature technologies can achieve  a physical 
  error lower than less mature technologies.}
\end{table*}

In addition to expanding the reach of traditional repeater networks, our mechanism may be used to hybridise
networks with different operating regimes. For example, quantum memories may be used to interconvert between
repeater networks with different networking protocols, qubit implementations, or operating rates \cite{S16}.  Quantum
memories embedded in portable devices may allow mobile nodes to connect to static networks.

%\section{Conclusions}

Our assumptions regarding the amount of space within each unit reserved for the actual qubits and necessary control infrastructure are somewhat arbitrary due to the unknown nature of how classical control technology will develop.  An underlying assumption of this work is that a high speed, high fidelity quantum network is only describable in a future where large-scale quantum technology exists.  While research related to the classical control of active quantum memories is in its infancy \cite{D14,F15,DFTMN10}, initial efforts in commercialising quantum technology suggests that our estimates are not overly optimistic.  D-WAVE provides an example of a commercial package that requires cryogenic cooling, power and classical control [The DWAVE two system (www.dwavesys.com/d-wave-two-system) utilises a closed cycle dilution refrigerator, magnetic shielding, vacuum system and classical control in a size approximately the same as a TEU container]. 

As quantum technology advances, a network architecture based on the transport of reliable quantum memories
could enable fundamental tests of quantum mechanics over long distances and then increasingly
sophisticated applications ranging from quantum cryptography to distributed quantum computing. Our mechanism
has the flexibility to service these applications as they develop, in addition to complementing traditional
repeater networks as they are deployed.  Eventually, once quantum computers are commonplace, entanglement will
be the fungible resource that enables a vast range of distributed applications. The quantum sneakernet is 
a mechanism that could feasibly underpin an entanglement-based economy of this kind, connecting users of
local quantum networks to a global Quantum Internet.

\section{Methods}

\begin{figure*}[ht!]
\begin{center}
\resizebox{170mm}{!}{\includegraphics{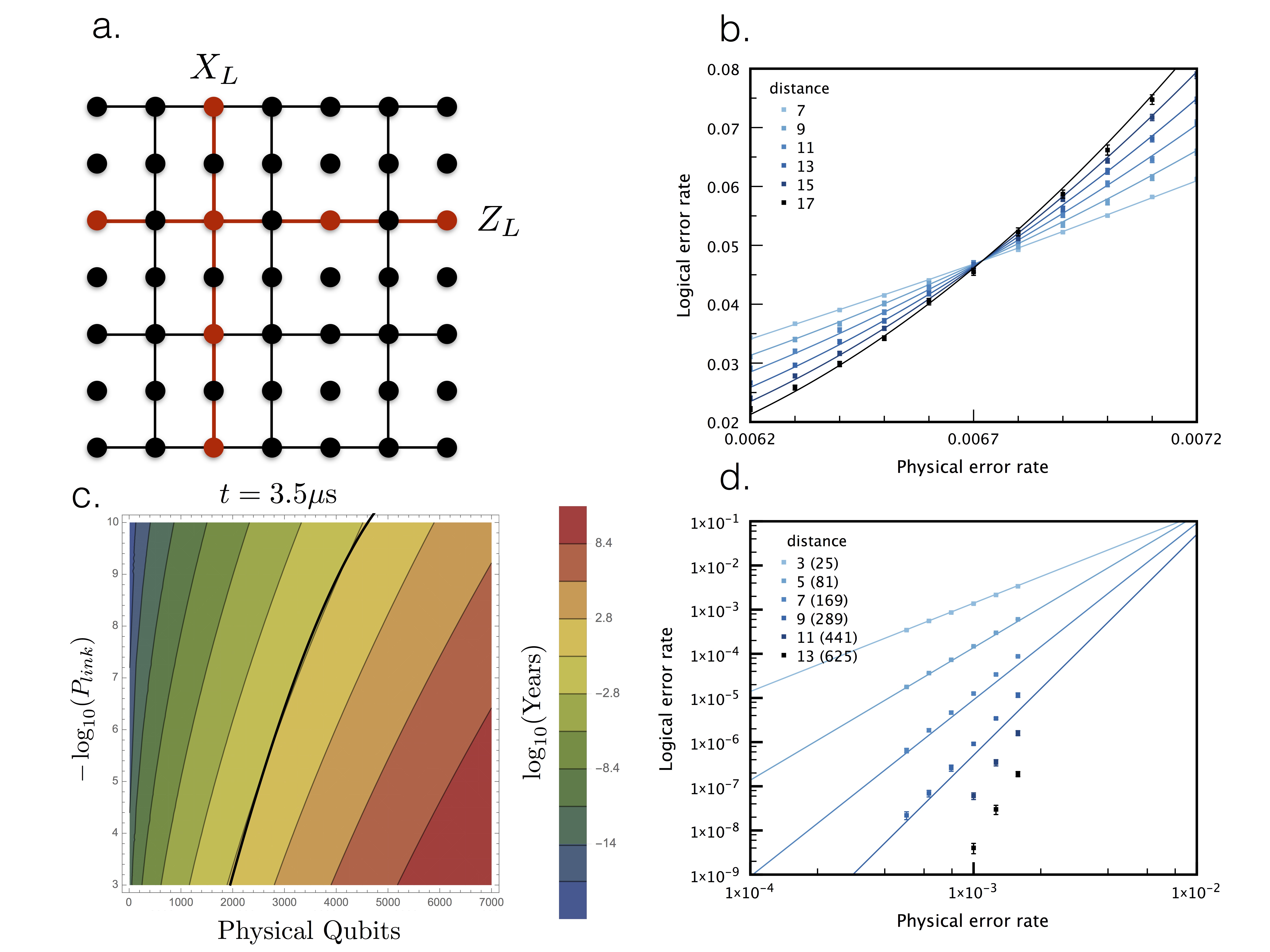}}
\end{center}
\vspace*{-15pt}
\caption{{\bf Properties of the planar code.} The memory unit is a two-dimensional nearest neighbour array of
  qubits encoded with the topological planar code \cite{DKLP02}.  {\bf a}, Structure of the planar code with
  the single qubit $X_L$ and $Z_L$ operators. (Gridlines do not represent entanglement bonds.)  {\bf b},
  Logical error rate as a function of physical error rate for different code distances.  Simulations confirm
  that the fault-tolerant threshold of the planar code lies at $p \approx 0.7\%$ under a 
  standard error model.  {\bf c}, Memory time as a function of $N$ and $P_{link}$ for a device that has a
  $t=3.5\mu$s physical gate time and $p=0.1\%$.  Along the heavy black contour line, an $N$-qubit memory unit can
  maintain sufficiently high coherence to achieve the selected $P_{\text{link}}$ after one year of storage
  time.  {\bf d}, Per-error correction cycle logical error rate $P_L$ as a function 
  of physical error rate $p$ for different code distances.  Numbers in
  parentheses are total physical qubits per logical qubit at that code distance.
%  At a physical error rate $p=0.1\%$, we can estimate the logical failure rate,
%  $P_L$, as a function of code distance, $d$.
}
\label{fig:memorystick}
\end{figure*}

{\bf Capacity of a memory unit: }Illustrated in Figure \ref{fig:memorystick} is the structure and performance of a memory unit.  The device
encodes a single logical qubit of memory [Figure \ref{fig:memorystick}{\bf a.}].  The logical Pauli operators
are chains of physical $X$ and $Z$ operations that connect the top and bottom (logical $X$) or the left and
right (logical $Z$) edges of the lattice.  Through simulation, we numerically determine both the
fault-tolerant threshold for the memory unit [Figure \ref{fig:memorystick}{\bf b.}]  and the expected failure
rate as a function of QEC strength at a fixed physical error rate, $p$ [Figure \ref{fig:memorystick}{\bf d.}].  From the behaviour of the code for
low values of the physical error rate, $p$, we can estimate the probability that a memory unit fails during
one error correction cycle, $P_L$, as a function of the distance of the topological planar code, $d$ (an error 
correction cycle requires $d$ rounds of error correction).  For an
operational device, we assume the error rate for each physical gate in the quantum memory is, $p$.  
The functional form for the failure of the code is given by, 
\begin{equation}
P_L \approx \alpha (\beta p)^{\frac{d+1}{2}}
\end{equation}
 We use the data from Figure \ref{fig:memorystick}, which simulates a full $d$ rounds of error correction 
 to estimate $\alpha \approx 0.3$ and $\beta \approx 70$.  The total
number of physical qubits in the memory unit is $N = (2d-1)^2$ and the total time of a memory correction cycle
is $T_{\text{corr}} = 6t d$, where $t$ is the operational time of a \emph{physical} quantum gate
(initialisation, measurement or \textsc{cnot}), the factor of 6 comes from the six elementary gates necessary
to perform syndrome extraction in the topological code, and we require $d$ rounds of error correction to
correct for measurement errors.

The total memory time of the unit, $T_{\text{mem}}$, is related to the per-error correction cycle failure
probability, $P_L$, and the chosen permissible infidelity of the final entangled link, $P_{\text{link}} =
1-F$, where $F$ is the link fidelity (between memory units),
\begin{equation}
\begin{aligned}
T_{\text{mem}} &= \frac{\log(1-P_{\text{link}})T_{\text{corr}}}{\log(1-P_L)} \\
&\approx \frac{6t(\sqrt{N}+1)P_{
\text{link}}}{2\alpha}(\beta p)^{-(\frac{\sqrt{N}+3}{4})}
\end{aligned}
\label{eq:mem}
\end{equation}
Figure \ref{fig:memorystick}{\bf b}. shows the memory time for a device that has a $t=3.5\mu$s physical gate
time (appropriate for optically coupled NV$^-$ \cite{N14}), as a function of total number of physical qubits, $N$, and desired final link infidelity,
$P_{\text{link}}$.  The contour where the memory unit can maintain coherence for one year and achieve the
desired link fidelity is plotted with a heavy line.  Similar plots can be easily obtained from
Eqn. \ref{eq:mem} for different physical gate times, $t$.  As it is assumed that the physical system is at a fixed physical error rate $p=0.1$\% (or $p=0.001$\%)
regardless of the intrinsic gate speed of the system, $t$, memory times will increase with slower systems.
Ion trap computers will have a {\em longer} memory time than donor-based systems as we assume {\em both}
technologies can achieve a $p=0.1\%$ (0.001\%) error rate on all fundamental gates.  
For a 
$N=4225$ qubit memory unit for the optically coupled NV$^-$ system.  
Taking $T_{\text{mem}} \approx 40$ days as our target memory time, we find the link
infidelity achievable is approximately $P_{\text{link}} \approx 10^{-10}$ (We assume a 40 day storage time for a 20 day transit time to account for preparing and consuming the Bell states at the source or destination.  This in practice can be 
reduced to 20 days by strategic choices of which Bell states to prepare and consume at given points in time, but this does not significantly change the size of each memorystick).  For all other technologies we 
recalculate the size of the memory unit to achieve the same infidelity and memory time given the 
physical gate time, $t$, and the physical error rate, $p$.

\textbf{Lattice surgery operations.} The planar code (and all toric code derivatives) allows a logical
two-qubit \textsc{cnot} gate to be executed as a transversal operation using individual \textsc{cnot} gates
applied between corresponding qubits in each memory unit.  While fault tolerant, this method may be difficult
to implement due to the difficulty of ensuring that each qubit in the 2D memory cell can interact
with the corresponding qubit in another cell.  A different approach, called \emph{lattice surgery}, partially
solves this problem by realising a fault-tolerant \textsc{cnot} gate between two memory units by only using
interaction between qubits along an edge of each memory unit.

Lattice surgery works by merging two separate lattices, each containing a single logical qubit encoded in the planar
code, into a single oblong lattice, then splitting up this single planar code again.  The merging operation is done by
matching the edges of two distinct logical qubits and measuring code stabilisers spanning the lattice cells.  This
effectively reduces a two qubit encoded system to a single encoded qubit.  This merging takes the state
$\ket{\psi}_L\otimes \ket{\phi}_L = (\alpha\ket{0}_L+\beta\ket{1}_L)\otimes (\alpha'\ket{0}_L+\beta'\ket{1}_L)$ to
$\alpha\ket{\phi}_L+\beta\ket{\bar{\phi}}_L=\alpha'\ket{\psi}_L+\beta'\ket{\bar{\psi}}_L$, where $\ket{\bar{A}} =
\sigma_x\ket{A}$.  The measurement of the stabilisers to perform a merge must occur $d$ times, where $d$ is the
effective code distance of each planar code.  This protects against faulty qubit measurements for each stabiliser
measurement.  Given that the quantum circuit required to measure the stabilisers for the planar code requires 6 physical
gates, the merge operation requires a time of $T=6td$, for physical gate times $t$.  
%{\color{blue}132}

The splitting operation is executed by physically measuring the qubits along the merged edge to divide the single lattice
back into two individual lattices.  The effect of a split operation is to take the single logical state encoded in the
joint lattice, $\alpha\ket{0}_L+\beta\ket{1}_L$ to the two-qubit state, $\alpha\ket{00}_L+\beta\ket{11}_L$.  Once again
to protect against measurement errors, error correction of both lattices must be run for a total of $d$ cycles,
requiring a total time of $T=6dt$ for the split operation.  
%{\color{blue}78}
 
Given these transformations, we can construct a Bell state between two encoded memory units by initialising a $d\times
d$ lattice holding a logical qubit in one memory unit in the $\ket{+}_L$ state and a logical qubit in the other memory
unit in the $\ket{0}_L$ state, merge the edges of the lattices across the optical interface between units to form a
single state $\ket{+}_L$ in a $2d\times d$ distance lattice, and then split them again to create the state
$\left(\ket{00}_L+\ket{11}_L\right)/\sqrt{2}$, with one logical qubit held in each memory unit.  This state can be
manipulated through transversal Hadamard operations on each memory cell and/or $X_L$ and $Z_L$ to any of the three other
Bell states in either the $X$- or $Z$-basis.  The total time for the split/merge operation will be $T=12dt= 6(\sqrt{N}+1)t$ for a
physical gate time of $t$ and a memory cell containing $N$ qubits.  
For the NV$^-$ design described above, $t=3.5\mu$s and for $N=4225$, $T\approx 1.4$ms.
%{\color{blue}80}

\textbf{Network operational procedures.}  Our network protocol ensures
that the effective transoceanic bandwidth is limited only by the
freight capacity and the transport time.  A total of seven shipping
containers are utilised for each ``online'' pair.  Two units are
permanently located at each shipping terminal.  Three mobile units
rotate locations; at any point in time, one is at each terminal ($A$
or $B$) and the third is aboard ship.  The protocol is separated into
three phases, which operate sequentially in each direction. In the
first phase, each mobile logical qubit is entangled with a stationary
logical qubit at the origin to establish logical Bell pairs.  In the
second phase, one logical qubit in each logical Bell pair is
transported from the origin to the terminus, undergoing continuous
error correction.  In the third phase, the logical qubits in the
logical Bell pairs are entangled with additional logical qubits at the
origin and the terminus and then measured to provide end-to-end
entanglement swapping at the logical level.

For example, consider the case in which a stationary unit sitting at
terminal $A$ is entangled with a mobile memory unit at terminal $B$,
and this pair is used as the online pair for supplying
terminal-to-terminal entanglement to other parts of the network.
After the remote entanglement supply in the mobile unit is exhausted,
it will be re-entangled with another stationary memory unit at its
current location.  A second mobile memory unit is aboard ship,
entangled with a stationary memory unit at the shipping terminal from
which it departed, carrying entanglement from $B$ to $A$.  A third
mobile memory unit at $A$ is creating entanglement with a stationary
local partner in preparation for shipping.  This ensures that ships
are never transporting inactive (unentangled) memory units.

The fixed constraints on the bandwidth of a link are latency of the
ship, the capacity of a memory unit, and the physical gate time, from
which we can derive additional operational procedures and hardware
development goals.  The fixed 20-day transit time for the
Japan-U.S. link serves as an upper bound for completing the
entanglement of a mobile memory unit with a stationary memory unit,
and for consuming the entanglement after shipping.  The $T = 1.4$ms
logical Bell pair creation time above arises from a surface code
distance $d = 33$ and gate operation time $t = 3.5\mu$s, for an NV$^-$
optical implementation \cite{N14}, and assumes that inter-container
operations can be executed at the same rate as operations local to
each memory unit.  (For the optical NV$^-$ system, this is a
reasonable assumption, because the intra-container operations use the
same physical mechanism as the inter-container operations.)  At this
operation rate, the entanglement of a memorystick containing
approximately 12.7KEb (Kilo-Entangled-bit) in the NV$^-$ optical
approach is created or consumed in 18s, and at full rate an entire
shipload of entanglement would be consumed in a little over two days.
Thus, inter-container operations may be $10\times$ slower without
impacting the performance even if only one container at a time out of
an entire shipload is used online.  To avoid long periods of
unavailability of the network, it may be desirable to limit the rate
at which entanglement is provided to applications to one-tenth of the
achievable physical rate.

The denser memory subsystems, differing physical gate times, and
varying code distances for other options in Table I
will result in different demands on the inter-container interfaces.
Because of the generic nature of the created entanglement, slow
inter-container interfaces can be compensated for by having more than
a single container online.

To achieve the performance in the right hand column of Table I, the
containers on board ship must collectively provide that much
bandwidth.  For the first three entries in the table, we expect that
having a single container online with a simple serial interface will
be sufficient.  For the higher data rates, parallelism can be used
both at the container interface and by having all $10^4$ containers
online at the same time.  Each container must provide $10^6-10^8$
logical entanglements/second.  These entanglement operations may be
multiplexed through a single connection, or more likely carried in
parallel through a parallel waveguide bundle.  Even so, achieving this
level of performance will require advances in optical entanglement
methods for the physical technologies.  Thus, this table serves as a
performance target for experimental work.

\section*{Acknowledgements} 

SJD acknowledges support from the JSPS Grant-in-aid for Challenging Exploratory Research and 
NICT, Japan.  RV and SJD acknowledge support from JSPS KAKENHI Kiban B 25280034.  RV acknowledges that this project has been made possible in part by a gift from the Cisco University Research Program Fund, a corporate advised fund of Silicon Valley Community Foundation.  ADG acknowledges the ARC for financial support 
(DP130104381).  AMS acknowledges support from NICT, Japan. 

\section*{Author contributions}

SJD and ADG conceived the idea.  AMS undertook the numerical simulations.  RV devised the network protocol.  All authors contributed to the writing of the manuscript.

%\section*{Additional information}

%Supplementary Information accompanies this paper. {\color{red} ??}

\section*{Competing financial interests}
The authors declare no competing financial interests.

%\section*{Data and materials availability: }
%All data needed to evaluate the conclusions in the paper are present in the paper and Supplementary Materials. Further information pertaining to the theoretical calculations reported in this work will be made available by the authors upon request.

\bibliographystyle{unsrt}

\end{document}